\documentstyle[12pt]{article}                    
\begin{document}    
\baselineskip=5mm     
\newcount\sectionnumber
\sectionnumber=0      
\pagestyle{empty}       
\begin{flushright}{UTPT--96--6}
\end{flushright}
\vspace{8mm}

\begin{center}  {\bf {\huge Charm and Bottom Semileptonic Decays } }\\  
\vspace{6mm}  Patrick J.
O'Donnell and  G\"{u}rsevil  Turan  \\  
Physics  Department,\\  
University  of
Toronto,\\ 
Toronto, Ontario M5S 1A7, Canada.  
\end{center}

\vskip 20pt \centerline{\bf Abstract}

We review the present status of theoretical attempts to calculate
the  semileptonic  charm and  bottom  decays  and then  present a
calculation  of these  decays  in the  light--front  frame at the
kinematic  point  $q^2=0$.  This  allows us to evaluate  the form
factors  at the same  value of $q^2$,  even  though  the  allowed
kinematic  ranges for charm and bottom decays are very different.
Also, at this kinematic point the decay is given in terms of only
one form  factor  $A_{0}(0)$.  For the  ratio of the decay  rates
given by the E653 collaboration we show that the determination of
the  ratio  of  the   Cabibbo--Kobayashi--Maskawa   (CKM)  matrix
elements is  consistent  with that  obtained  from the  unitarity
constraint.  At present,  though, the unitarity  method still has
greater accuracy.  Since  comparisons of the semileptonic  decays
into $\rho$ and either  electrons or muons will be available soon
from the E791  Fermilab  experiment,  we also look at the massive
muon  case.  We show  that for a range  of  $q^2$  the  $SU(3)_F$
symmetry  breaking is small even though the  contributions of the
various helicity  amplitudes  becomes more  complicated.  For $B$
decays,  the  decay $B  \rightarrow  K^{*}  \ell  \bar{\ell}$  at
$q^2=0$  involves  an extra form  factor  coming  from the photon
contribution and so is not amenable to the same kind of analysis,
leaving only the decay $B  \rightarrow  K^{*}\nu  \bar{\nu}$ as a
possibility.  As the mass of the decaying  particle  increases we
note that the $SU(3)$ symmetry becomes badly broken at $q^2=0$.

\newpage \pagestyle{plain}

\section{Introduction}

Semileptonic B- and D- meson decays constitute a subject of great
interest in the physics of electroweak  interactions, as they may
help determine the various CKM mixing angles.  In particular, the
decays involving  $b\rightarrow  c\ell  \nu_{\ell}$ are eminently
suitable for the heavy quark effective theory (HQET) to determine
\cite{heavy} the CKM matrix  $V_{cb}$.  For exclusive decays to a
final state with a $u$ or $s$ quark, and for the D--meson  decays
as a whole, it is less  likely  that the heavy  quark  symmetries
apply.  Since  the   dynamical   content  of  the   corresponding
amplitudes is contained in Lorentz-  invariant  form  factors, to
know and  understand  form  factors of hadronic  currents is very
important for analyzing these decays.

However,   few  of  these  form   factors   have  been   measured
experimentally,  and those  that have  been, are not  known  very
precisely  yet because of the  smallness of the branching  ratios
associated  with  them.  On  the  other  hand,  the   theoretical
calculations are hard to estimate because of the  nonperturbative
character  of strong  interactions.  Here,  one may  resort  to a
model, but that  introduces  uncertainties  that are inherent to
the model itself.  To overcome this  difficulty, at least to some
degree,  many  authors  have  studied,  instead of the  branching
ratios of the semileptonic decays of the particular heavy mesons,
their ratios at some particular  kinematical  points,  usually at
zero recoil.

For example, Sanda and Yamada \cite{SY} propose a strategy to get
$|V_{ub}|$  by  relating  the  differential  decay  width  of the
$\bar{B}  \rightarrow  \rho \, \ell \, \bar{\nu}$  to that of the
process  $\bar{B}  \rightarrow  K^{*} \, \ell \,  \bar{\ell }$ at
their   respective    $q^{2}_{max}$   limits   ($(m_{\bar{B}}   -
m_{\rho})^2$  and   $(m_{\bar{B}}  -  m_{K^{*}})^2$)   using  the
$SU(3)$--flavor  symmetry  and heavy  quark  approximation.  They
find

\begin{equation} 
\frac{|V_{ub}|^{2}}{|V_{tb}V^{*}_{ts}|^{2}}=
\frac{q^{2\bar{B}\rightarrow       K^{*}}_{max}}{q^{2
\bar{B}\rightarrow \rho}_{max}}  \left(\frac{p_{K^{*}}}{p_{\rho}}
\right)_{lim}   \left(\frac{\alpha_{QED}}   {4  \pi}  \right)^{2}
2(C^{2}_{V}+C^{2}_{A})  \frac{[d\Gamma  (\bar{B}\rightarrow  \rho
\ell        \bar{\nu}        )/dq^{2}]_{q^{2}         \rightarrow
q^{2\bar{B}\rightarrow       \rho       }_{max}}}       {[d\Gamma
(\bar{B}\rightarrow   K^{*}  \ell   \bar{\ell}   )/dq^{2}]_{q^{2}
\rightarrow  q^{2\bar{B}\rightarrow K^{*} }_{max}}}  
\label{eqSY17} 
\end{equation}
where $(p_{K^{*}}/p_{\rho})_{lim}=\sqrt{m_{\rho}/m_{K^{*}} }$ and
$C_{V}$ and $C_{A}$ are the QCD  corrected  Wilson  coefficients.
The matrix element $|V_{ub}|$ may be determined if the RHS can be
obtained  by  experiment  and   $|V_{ts}|$   from  the  unitarity
condition.  The problem is that in the zero recoil limit $p_{\rho
, K^{*}}\rightarrow 0 $, i.e.,  $q^{2}=q^{2}_{max}$,  the $q^{2}$
distributions  vanish due to the phase  space  suppression.  This
means that experimentally there should be no events at that point
and  very few in the  neighborhood,  making  it a very  difficult
measurement.

Dib and Vera  \cite{DV}  relate  $B  \rightarrow  \rho \, \ell \,
\bar{\nu}$ to $ D \rightarrow  \rho \, \ell \, \bar{\nu}$ also at
the point of zero recoil, using the heavy quark  symmetry, to get
a model independent  result to leading order in inverse powers of
large masses.  At the  kinematical  point of zero recoil,  $y=1$,
where $y=(m^{2}_{I} + m^{2}_{\rho} - q^{2})/2m_{I}  m_{\rho}$ and
$I=B$ or $D$ they find

\begin{eqnarray}  
\frac{d\Gamma  (B\rightarrow  \rho e \nu  )/dy}
{d\Gamma (D \rightarrow  \rho e \nu )/dy}  \begin{array}{l}  \mid
\\[-0.3cm]  \mid  _{y\rightarrow  1} \end{array} & = & \frac{\mid
V_{ub}\mid^{2}}{\mid            V_{cd}\mid^{2}}            \left(
\frac{f^{[B]}_{A_{1}}(1)}{f^{[D]}_
{A_{1}}(1)}\right)^{2}\left(\frac{m_{B}-m_{\rho}}{m_{D}-m_{\rho}}
\right)^{2} \label{eqDV6} 
\end{eqnarray}

They      determine     the     ratio     of     form     factors
$f^{[B]}_{A_{1}}(1)/f^{[D]}_{A_{1}}(1)$  in the constituent quark
model at the  tree-level  in the HQET and also with the inclusion
of  short--distance  QCD corrections.  Their numerical result for
the  ratio  of  form  factors  is  between  1.09  and  1.18.  The
parameters  that cause the largest  uncertainty  in the ratio are
$m_{c}$  and $\mu $.  Again,  the rates  vanish  at $y=1$  due to
phase   space.   To   determine   the   LHS   of    (\ref{eqDV6})
experimentally   one  should   access   the  region   nearby  and
extrapolate to the point of $y=1$.

Ligeti and Wise give a  model-independent  method \cite{LW} which
is based  on the  study  \cite{bg}  of the  double  ratio of form
factors      $(f^{(B\rightarrow\rho      )}/     f^{(B\rightarrow
K^{*})})/(f^{(D\rightarrow\rho   )}/f^{(D\rightarrow   K^{*})})$.
They  claim  that  this  double  ratio  is equal to  unity in the
$SU(3)$ limit, and in the limit of heavy quark symmetry so that a
determination   of  $|V_{ub}|$  is  possible  using   information
obtainable   from  the  decay  modes   $B\rightarrow   \rho  \ell
\bar{\nu}_{\ell}$,    $B\rightarrow    K^{*}   \nu    \bar{\nu}$,
$D\rightarrow  \rho  \bar{\ell }  \nu_{\ell}$  and  $D\rightarrow
K^{*} \bar{\ell}  \nu_{\ell}$.  They use a pole model to get away
from the  zero--recoil  point.  Since the maximum  values for $y$
are  different  for  $B$  and  $D$  decays,   $y_{max}=3.5$   and
$y_{max}=1.3$, respectively, they limit $y$ for the $B$ decays to
lie in the  range  $1< y  <1.5$.  Provided  $f^{(D\rightarrow\rho
)}(y)/f^{(D\rightarrow  K^{*})}(y)$ is almost  independent of $y$
then a precise  value for  $|V_{ub}|$  can be extracted  from the
rates for  $B\rightarrow  K^{*} \nu \bar{\nu}$ and  $B\rightarrow
\rho \ell  \bar{\nu}_{\ell}$  integrated  over this region in $y$
and  $f^{(D\rightarrow\rho   )}(1)/f^{(D\rightarrow  K^{*})}(1)$.
However, at the present time the rare decays  $B\rightarrow K^{*}
\bar{\ell  }   \nu_{\ell}$   or  and   $B\rightarrow   K^{*}  \nu
\bar{\nu}$,  have not been observed, and there is no  information
on the individual form-factors for $D\rightarrow \rho \bar{\ell }
\nu_{\ell}$.

In this paper, we  concentrate on opposite end of the heavy meson
decay  kinematic   spectrum,   namely,  vanishing   four-momentum
transfer,  $q^{2}=0$,  for the  ratio  of  $D\rightarrow  \rho \,
\bar{\ell} \, \nu_{\ell}$ to $D\rightarrow K^{*} \, \bar{\ell} \,
\nu_{\ell}$  and for the ratio of  $B\rightarrow  \rho \, \ell \,
\bar{\nu_{\ell}}$  to  $B\rightarrow  K^{*} \, \nu \, \bar{\nu}$.
This  kinematic  point is the  maximum  recoil of the  $\rho$  or
$K^{*}$; $q^2=0$  corresponds to different values of $y$ in these
cases and is not  therefore  a good point  from the $y$, or heavy
quark approach.  The motivation for choosing this kinematic point
is  that  first  of all,  there  is a well  developed  way in the
light-front  formalism  \cite{model}  to deal  with the  point at
$q^{2}=0$.  Secondly, the other  calculations  obtain  results at
the zero  recoil  point  where it is known  that the  experiments
should find no events so that extrapolations and pole models have
to be used.  Thirdly,  the decay  widths at $q^2=0$  are given in
terms  of  only  one  form  factor  $A_{0}(0)$  (defined  below).
Finally,  there is now a first report of the lattice  calculation
of    the    form    factor     $A_{0}(0)$    for    the    decay
$\bar{B^{0}}\rightarrow   \rho^{+}   \ell^{-}   \bar{\nu_{\ell}}$
\cite{UKQCD},  which  is  important  phenomenologically  for  the
determination  of  $|V_{ub}|$.  They have  determined  a range of
values for $A_{0}^{B^{0}\rightarrow  \rho^{+}}(0)$ and found that
$A_{0}^{B^{+}      \rightarrow       \rho^{0}}(0)/\sqrt{2}      =
(0.16-0.35)^{+9}_{  \!  \,  -6}$,  where  the  range  is  due  to
systematic uncertainty and the quoted error is statistical.

\section{Semileptonic   $D\rightarrow  V  \ell  \nu_{\ell}$
Decays}

We  define  the  form  factors  in the  semileptonic  decay  of a
$D$-meson  $D(c\bar{q})$ into a vector meson  $V(Q\bar{q})$ with
the    polarization    vector     $\varepsilon^{\mu}    $,    by
 
\begin{eqnarray}
<V(p_{V},\varepsilon  )|\bar{Q}  \gamma_{\mu } (1-\gamma_{5} )c |
D(p_{D})> & = & \frac{2V(q^{2})}{m_{D}+m_{V}}  i \varepsilon_{\mu
\nu        \alpha        \beta        }        \varepsilon^{*\nu}
p^{\alpha}_{D}p^{\beta}_{V}-(m_{D}+m_{V})\varepsilon^{*\mu}
A_{1}(q^{2})      \nonumber     \\     &     &      \hspace{-3cm}
+\frac{(\varepsilon^{*}                                     \cdot
p_{D})}{m_{D}+m_{V}}(p_{D}+p_{V})^{\mu}       A_{2}(q^{2})-2m_{V}
\frac{(\varepsilon^{*}   \cdot  p_{D})}{q^{2}}  q^{\mu}  A(q^{2})
\end{eqnarray}   
The  form   factor   $A$  can  be   written   as
\begin{eqnarray}   A(q^{2})   &  =  &   A_{0}(q^{2})-A_{3}(q^{2})
\end{eqnarray}   
where   
\begin{eqnarray}   A_{3}(q^{2})  &  =  &
\frac{m_{D}+m_{V}}{2m_{V}}                          A_{1}(q^{2})-
\frac{m_{D}-m_{V}}{2m_{V}}   A_{2}(q^{2})   \label{def}
\end{eqnarray}   
and with $A_{0}(0)=A_{3}(0)$.

In the limit of vanishing lepton masses, the term proportional to $A$
in eq.(3) does not contribute to the total amplitude and hence to
the decay rate. In this limit, the differential $q^{2}$-distribution
of the semileptonic  $D\rightarrow V \ell \nu_{\ell}$ decay can be written  
\begin{eqnarray}
\frac{d\Gamma}{dq^{2}}|_{m_{\ell}=0} & = &  \frac{G^{2}}{(2\pi)^{3}} 
|V_{cQ}|^{2} \frac{p_{V} q^{2}}{12 m_{D}^{2}} 
 (|H_{+}|^{2}+|H_{-}|^{2}+|H_{0}|^{2}) \label{dg1}
\end{eqnarray}
where   $H_{+},H_{-}$   and  $H_{0}$  are  the  partial  helicity
amplitudes, 
\begin{eqnarray}
H_{\pm} & = & 2m_{V}A_{0}(q^{2})+ 
\frac{(m^{2}_{D}-m^{2}_{V}-q^{2})A_{2}(q^{2})\mp 
2m_{D}p_{V} V(q^{2})}{m_{D}+m_{V}} \nonumber \\ 
H_{0} & = & \frac{1}{\sqrt{q^{2}}} \left[ (m^{2}_{D}-m^{2}_{V}-q^{2})
A_{0}(q^{2})+\frac{2m_{V}q^{2}}{m_{D}+m_{V}}A_{2}(q^{2}) \right] \; .
\end{eqnarray}

We now compare the lepton  spectra in the decays  $D\rightarrow  \rho
\bar{\ell}   \nu_{\ell}$  and   $D\rightarrow   K^{*}  \bar{\ell}
\nu_{\ell}$  at $q^2=0$.  In the limit of vanishing  lepton mass,
the differential  decay rate for $D\rightarrow V \ell \nu_{\ell}$
decay is determined by only one form factor $A_{0}$ at $q^{2}=0$:

\begin{eqnarray}     
\frac{d\Gamma(D\rightarrow V \ell \nu_{\ell})}{d q^{2}} \mid_{q^{2}\rightarrow0} 
& = & \frac{G^{2}_{F}}{192 \, \pi^{3} \, m^{3}_{D}} \, |V_{cQ}|^{2}
\,  (m^{2}_{D}-m^{2}_{V})^{3} \, |A^{D\rightarrow  V}_{0}(0)|^{2} 
\label{eq:OT0}   
\end{eqnarray}  

Hence,  the  ratio  of  the  two distributions  at $q^{2}=0$  is  
\begin{eqnarray}  
\frac{[d\Gamma(D\rightarrow \rho \bar{\ell }\nu_{\ell})/dq^{2}]_{q^{2}
\rightarrow    0}}    {[d\Gamma (D\rightarrow  K^{*} \bar{\ell}
\nu_{\ell}   )/dq^{2}]_{q^{2}   \rightarrow   0}}   &  =  &
\frac{|V_{cd}|^{2}}{|V_{cs}|^{2}} \left(\frac{m^{2}_{D}-
m^{2}_{\rho}}{m^{2}_{D}-m^{2}_{K^{*}}}\right)^{3}
\frac{|A^{D\rightarrow \rho}_{0}(0)|^{2}}{|A^{D\rightarrow
K^{*}}_{0}(0)|^{2}} \label{eq:OT1}
\end{eqnarray}

From  the  experimental   ratio  $[d\Gamma   (D\rightarrow   \rho
\bar{\ell}  \nu_{\ell}  )/dq^{2}] /[d\Gamma  (D\rightarrow  K^{*}
\bar{\ell} \nu_{\ell} )/dq^{2}]_{q^{2}  \rightarrow 0}$ one needs
knowledge  of  $|A^{D\rightarrow   \rho}_{0}(0)|/|A^{D\rightarrow
K^{*}}_{0}(0)|$  to  extract  the ratio  $|V_{cd}/V_{cs}|$.  This
ratio is often  taken to have the value unity by  $SU(3)$--flavor
symmetry.

There have been many model--dependent studies of this ratio which
we show in table  \ref{tab1}.  From table \ref{tab1}, we see that
theoretical  predictions of the ratio of  form-factors  fall in a
range near $1$.  The procedure  adopted to get  $A_{0}(0)$  often
uses Eq.  (\ref{def})  with the calculated  values of $A_{1}$ and
$A_{2}$ rather than $A_{0}$ directly.  However, this indirect way
to get $A_{0}$ may have some difficulties coming from the $q^{2}$
dependences   of  the  form   factors  and  also  from   possible
correlations  in  treating  the  errors.  This has  already  been
commented on in ref.  \cite{CFS} and we will have further remarks
to make when we discuss  the  non-massless lepton case.  

We use the  light-front  quark  model,  which is  suitable at the
kinematic  limit where  $q^{2}=0$,  to determine  the same ratio.
This model was developed  \cite{model}  a long time ago and there
have been many applications  \cite{Jaus90}-\cite{GuHu}  where the
details can be found.

In the light -front quark model, the quark coordinates are given by
\begin{eqnarray} 
& & p_{Q+}=xP_{+} \;\; , \;\; 
{\bf p}_{Q\perp}=x{\bf P_{\perp}}+{\bf k}_{\perp} \;\; , \;\;
p_{\bar{q}+}=(1-x)P_{+} \;\; , \;\; 
{\bf p}_{\bar{q}\perp}=(1-x){\bf P_{\perp}}+{\bf k}_{\perp}, \nonumber \\
& & 0 \leq x \leq 1 \;\;\; , \;\;\; {\bf P}={\bf p}_{Q}+{\bf p}_{\bar{q}}.
\end{eqnarray}
where ${\bf k}=(k_{z},{\bf  k}_{\perp})$ is the internal momentum.
For ${\bf P}$ (and similarly for other vectors), ${\bf P}=(P_{+},
{\bf P_{\perp}})$ with $P_{+}=P_{0}+P_{z}$ and ${\bf P_{\perp}}=(P_{x},P_{y})$ .

To calculate  the form  factors,  one  reasonable  and often used
assumption for the meson wave function $\phi(x,{\bf  k}_{\perp})$
is a Gaussian-type function

\begin{eqnarray}
\phi(x,{\bf k}_{\perp}) & = & \eta({\bf k}) \sqrt{\frac{d{\bf k_{z}}}{dx}}
\;\;\; , \;\;\; \eta({\bf k})= (\pi \omega^{2})^{-3/4}
\exp{(-{\bf k}^{2}/2\omega^{2})} \label{wf1}
\end{eqnarray}
where $\omega $ is a scale  parameter and $x$ is defined through
\begin{eqnarray}
x & = & \frac{e_{Q}+k_{z}}{e_{Q}+e_{\bar{q}}} \;\;\; , \;\;\; 
e_{i}=\sqrt{m^{2}_{i}+{\bf k}^{2}} \;\; (i=Q,\bar{q}).
\end{eqnarray}

The   wave   function   (\ref{wf1})   has   been   used  in  ref.
\cite{model},    \cite{Jaus90}    and   also   in
\cite{OXT},\cite{ox1} for various applications of the light-front
quark model.

A similar wave function is 
\begin{eqnarray}
\phi(x,{\bf k}_{\perp}) & = & \eta({\bf k}) \sqrt{\frac{d{\bf k_{z}}}{dx}} 
\;\;\; , \;\;\; \eta({\bf k})= N
\exp{(-M^{2}_{0}/ 8 \omega^{2})} \label{wf2}
\end{eqnarray}
Here  $N$  is the  normalization  constant  and  $M_{0}$  is  the
invariant  mass of the quarks, which is now given by
\begin{eqnarray}
M_{0} & = & e_{Q}+e_{\bar{q}}
\end{eqnarray}
This wave function has been also
applied for heavy mesons in \cite{Jaus96} and \cite{GuHu}.

Another possibility is the wave function adopted in \cite{WSB}:
\begin{eqnarray}
\phi(x,{\bf k}_{\perp}) = N\sqrt{\frac{x(1-x)}{\pi \omega^{2}}}
\exp{\left(-\frac{M^{2}}{2\omega^{2}}
\left[x-\frac{1}{2}-\frac{m^{2}_{Q}-m^{2}_{\bar{q}}}{2M^{2}}\right]^{2} \right)}
\exp \left(-\frac{{\bf k}^{2}_{\perp}}{2\omega^{2}}\right) \label{wf3}
\end{eqnarray}
where $M$ is the mass of the meson.  As shown in \cite{OXT},  the
wave functions  (\ref{wf1})  and (\ref{wf3})  satisfy the scaling
law \cite{Neu}
\begin{eqnarray}
f_{H} & \propto & \frac{1}{\sqrt{m_{h}}} \;\;\; , \;\;\; m_{h} \rightarrow \infty 
\;\; , \label{slaw}
\end{eqnarray}
where $f_{H}$ is the heavy  $H$-meson  decay constant and $m_{h}$
is the  corresponding  heavy quark.  However,  the wave  function
(\ref{wf2})  does not satisfy  (\ref{slaw})  unless the parameter
$\omega$  scales  as the  square  root of the  heavy  quark  mass
\cite{privcom}.

We use the wave function  (\ref{wf1})  in our  calculations.  The
parameters   for  $\rho  $  and   $K^{*}$  are  taken  from  ref.
\cite{Jaus91}.  In ref.  \cite{Jaus91}  the pion  decay  constant
used has the  value  $f_{\pi}=92.4\pm  0.2 $ MeV and the  $\rho $
decay constant  $f_{\rho}/m_{\rho}=152.9 \pm 3.6$ MeV, both taken
from  experiment.  The same  value of  $\omega$  (called  $\beta$
there) is assumed for both the $\pi$ and the $\rho$ mesons.  This
is in line with the usual ideas of hyperfine splitting \cite{FO}.
The parameters  that are fitted are the quark masses, found to be
$m_{u}=m_{d}=250   \pm  5  $  MeV,  where  it  is  assumed   that
$\omega_{u\bar{u}}=\omega_{d\bar{d}}=\omega_{u\bar{d}} = 0.3194$.
A similar  calculation  for the kaon based on the decay  constant
$f_{K}=113.4   \pm   1.1  $  MeV   and   the   decay   rate   for
$K^{*+}\rightarrow  K^{+} \gamma $, leads to the $s$-quark mass $
m_{s}=0.37  \pm  0.02  $ MeV  and  the  wave  function  parameter
$\omega_{u\bar{s}}=0.3949$  GeV.  The  values  for the  masses of
$u$-  and  $s$-   quarks   and  the   wave   function   parameter
$\omega_{u\bar{s}}$ obtained in this way are used to evaluate the
$K^{*}$ decay constant, $f_{K^{*}}=186.73$ MeV.

Using  these  values  of  the  quark  masses  and  wave--function
parameter we get:  $A^{D\rightarrow \rho }_{0}(0)/A^{D\rightarrow
K^{*}}_{0}(0)=0.88$, i.e., an $SU(3)_{F}$--breaking effect at the
level   of   about   10   $\%$.   The   kinematical   factor   in
eq.(\ref{eq:OT1}) readjusts this value and the ratio of the decay
rates in terms of the CKM factor  becomes  $0.96$,  which is very
close to the  $SU(3)_{F}$  symmetry  limit.  This  result for the
values  of the form  factors  is not  strongly  dependent  on the
choice  of  wave  function.  We have  calculated  the  same  form
factors also with the wave function  (\ref{wf2})  above and got a
similar  result.  The fact that form factors do not depend on the
choice of wave  function can also be seen by comparing our result
with the one  given in the  first  row of the  table  \ref{tab1},
which came from using the wave function (\ref{wf3}).

The E653  Collaboration  determined the following  ratio of decay
rates \cite{E653}:  
\begin{eqnarray} \frac{\Gamma  (D^{+}\rightarrow \rho^{0}
\mu^{+} \nu )} {\Gamma (D^{+}\rightarrow  \bar{K^{*}}^{0} \mu^{+}
\nu )} & = &  0.044  ^{+0.031}_{\! \, -0.025} \pm 0.014 \label{eq:E653} 
\end{eqnarray}

Using this and the values  given in the last row of the Table for
the  form  factors  of  $D\rightarrow  \rho  $ and  $D\rightarrow
K^{*}$, we can extract the  following  result on the ratio of the
CKM   matrix    elements    $|V_{cd}/V_{cs}|$:   
\begin{eqnarray}
|V_{cd}|/|V_{cs}|  & = &  0.214  ^{+0.074}_{  \!  \,  -0.060} \; .  
\label{eq:OT3}
\end{eqnarray}  
which is consistent  with, but not as accurate as, the prediction
derived from the values quoted in PDG \cite{PDG}, coming from the
unitarity constraint:
\begin{eqnarray}
|V_{cd}|/|V_{cs}|  & = & 0.226  \pm  0.003 \; .  \label{eq:OT3.a}
\end{eqnarray}

\subsection{Lepton Mass Effects}

Most  of  the  theoretical  and  experimental   analyses  of  the
exclusive  semileptonic decays assume that taking the lepton mass
to  be  zero  is a  good  approximation.  For  the  electron  and
$\tau$--lepton cases, the situation is relatively clear.  For the
electron, the zero mass approximation is good since the threshold
is very close to the  massless  limit.  On the other  hand, it is
obvious  that  one  has  to  include  lepton  mass  effects  when
analysing  semileptonic decays involving $\tau $--leptons.  Since
comparisons  of the  semileptonic  decays into  $\rho$ and either
electrons or muons will be available  soon  \cite{Appel},  we now
discuss the massive muon case.

Two  different  aspects  have to be  considered  when lepton mass
effects  are  included  in an  analysis  of  semileptonic  decays
\cite{korner90}.  The kinematics of the decay  processes  change.
There is also a change of a  dynamical  nature:  when the  lepton
acquires a mass there can also be spin--flip contribution.

When  the  lepton  mass is  taken to be  nonzero,  Eq.(\ref{dg1})
becomes \cite{korner90}
\begin{eqnarray}
\frac{d\Gamma}{dq^2}|_{m_{\ell}\neq 0} & = & \frac{G^{2}}{(2\pi)^{3}}\, 
|V_{cQ}|^{2} \, \frac{p_{V} \, (q^{2}-m^{2}_{\ell})^{2}}{12 m_{D}^{2}
 q^{2}} \left[ (|H_{+}|^{2}+|H_{-}|^{2}+|H_{0}|^{2}) \right.\nonumber \\
& & \left. \hspace{2cm} +\left( \frac{m^{2}_{\ell}}{2q^{2}} \right)
(|H_{+}|^{2}+|H_{-}|^{2}+|H_{0}|^{2}+3|H_{t}|^{2})\right] \label{dg2}
\end{eqnarray}
In  addition  to spin 1  contributions,  there  are  off--shell spin 0
contributions proportional to $|H_{t}|^{2}$ where
\begin{eqnarray}
H_{t} & = & \frac{2m_{D}p_{V}}{\sqrt{q^{2}}} \left[ A_{0}(q^{2})
-\frac{q^{2}}{2m_{V}(m_{D}+m_{V})}A_{2}(q^{2}) \right] 
\end{eqnarray}

In   Eq.(\ref{dg2}),   spin  flip  contributions   bring  in  the
characteristic flip factor  $m^{2}_{\ell}/2q^{2}$  which vanishes
in the zero  lepton mass limit.  The bounds on $q^{2}$  are given
by  $m^{2}_{\ell}  \leq q^{2} \leq  (m_{D}-m_{V})^{2}$  and it is
seen  that  because  of  the  factor   $(q^{2}-m^{2}_{\ell})^{2}$
multiplying  all of the helicity  amplitudes, all the form factor
contributions vanish at threshold  $q^{2}=m^{2}_{\ell}$.  This is
in contrast to the case for  $m_{\ell}=0$,  Eq.(\ref{dg1})  where
the  longitudinal  helicity  amplitude  $H_{0}$  appears  with  a
$1/q^{2}$  factor which survives at $q^{2}=0$ with a contribution
proportional  to a single form factor  $A_{0}$.  We have  written
the helicity  amplitudes  in terms of the form  factors  $A_{0}$,
$A_{2}$  and  $V$  rather  than  the  more  conventional   choice
\cite{PDG,korner90}  of $A_{1}$,  $A_{2}$  and $V$.  This  latter
choice does not connect  smoothly to the massless  $q^2=0$  limit
and gives incorrect results for $SU(3)_F$ breaking.

To investigate the $SU(3)_{F}$  limit when the lepton mass effect
is included we need to consider  $q^{2}$  dependences of the form
factors.  For this we use the approximation
\begin{eqnarray}
F(q^{2}) & \simeq & \frac{F(0)}{1-q^{2}/\Lambda^{2}_{1}+q^{4}/
\Lambda^{4}_{2}}
\end{eqnarray}
where we take the  values  of the  parameters  $\Lambda_{1}$  and
$\Lambda_{2}$ from \cite{JausF}.

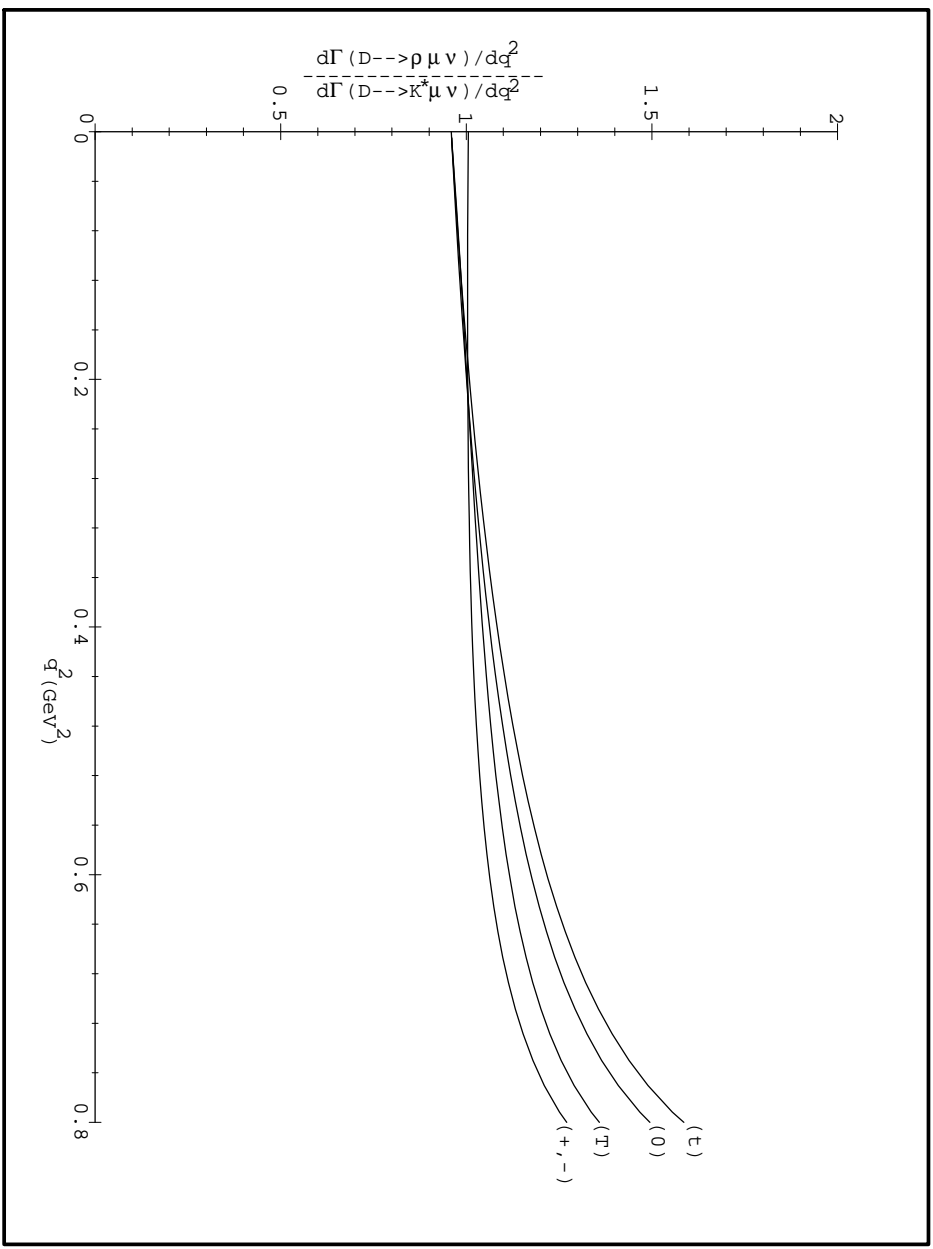
\begin{figure}[htb]
\caption{$q^{2}$ spectra for $[d\Gamma  (D\rightarrow \rho
\mu  \nu  )/dq^{2}]  /[d\Gamma   (D\rightarrow   K^{*}  \mu
\nu   )/dq^{2}]_{m_{\mu}\neq  0}$  for  different  helicity
contributions, longitudinal (0), tansverse $(+,-)$, flip (t) and total (T)
in units of $|V_{cd}/V_{cs}|^{2}$. \label{Fig1}}
\end{figure}

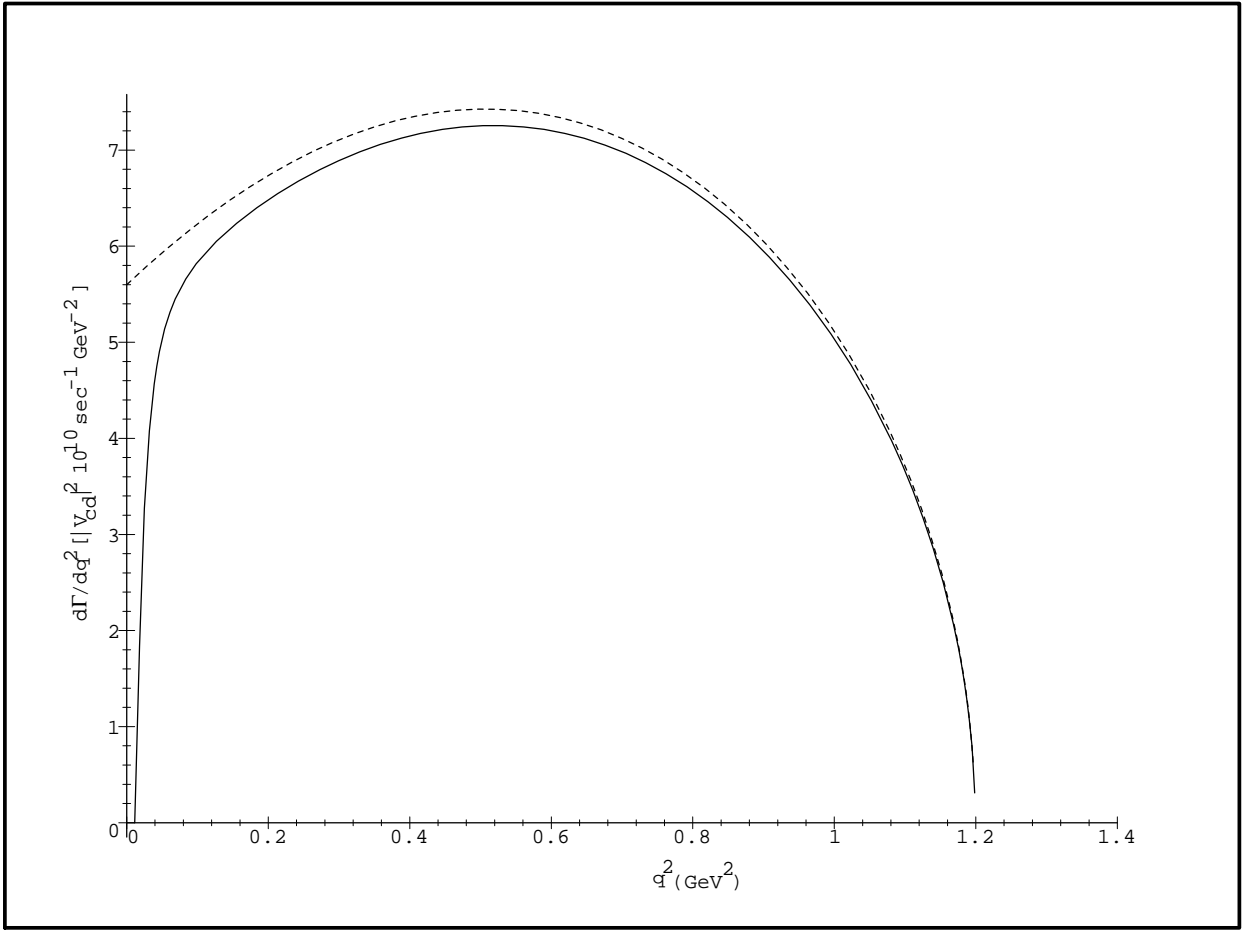
\begin{figure}[htb]
\caption{$q^{2}$ spectra of semileptonic  decay rates $D\rightarrow \rho
\mu \nu_{\mu}$  with $m_{\mu}=0$ (dotted) and $m_{\mu}\neq 0$ (full)
\label{Fig2}}
\end{figure}

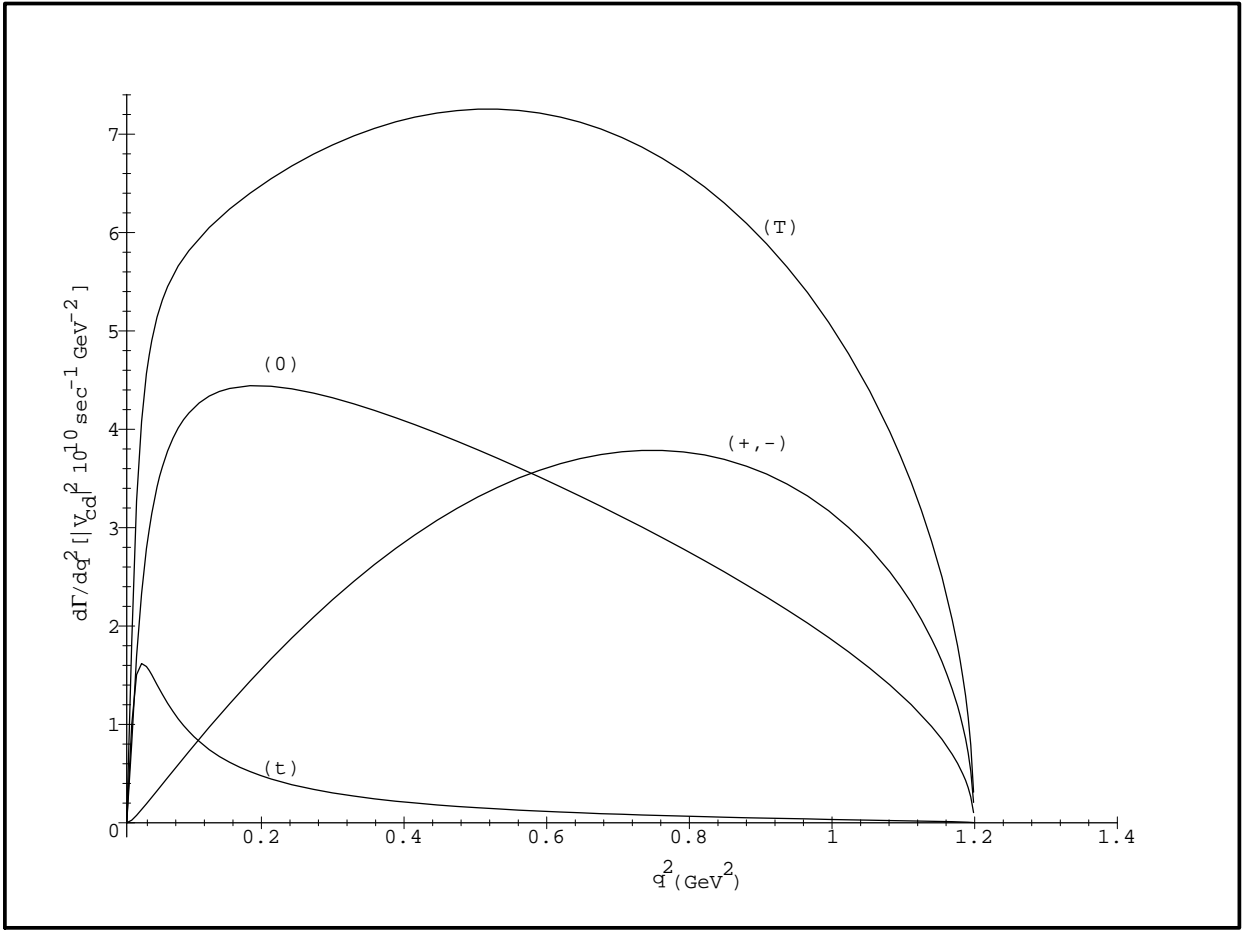
\begin{figure}[htb]
\caption{Partial helicity rates $d\Gamma/dq^{2}$ for  longitudinal (0), 
transverse (+,$-$), flip (t) and total (T) contributions 
as a function of $q^{2}$ for $D\rightarrow \rho \mu \nu_{\mu} $ 
($m_{\mu}\neq 0$). The flip contribution is small but not as tiny as 
indicated in ref.[21] \label{Fig3}}
\end{figure}

In Fig.  \ref{Fig1}  we show the  $q^{2}$  spectra  for the ratio
$[d\Gamma  (D\rightarrow \rho \mu \nu_{\mu}  )/dq^{2}]  /[d\Gamma
(D\rightarrow K^{*} \mu \nu_{\mu}  )/dq^{2}]_{m_{\mu}\neq 0}$ for
different helicity  contributions  defined in Eq.(\ref{dg2}).  As
an  explicit  example, if we take the mass of the muon to be zero
then  our  differential   decay  rate  for  $D\rightarrow   K^{*}
\mu\nu_{\mu}$  at  $q^{2}=0$  gives a value of $5.8$ in units  of
$|V_{cs}|^{2}10^{10}sec^{-1}GeV^{-2}$.    The    threshold    for
non-zero mass is  $q^{2}=0.011  \; GeV^{2}$  where the decay rate
vanishes.  It is at  $q^{2}=0.087  \; GeV^{2}$  that the value of
$5.8$ is first  obtained.  However, this $5.8$ is now composed of
three  parts  (see  table   \ref{tab2}),  that  coming  from  our
$A_{0}(q^2)$  with a value of $4.15$  and two  other  parts,  one
called a flip  contribution  (giving  $1.04$) with the  remaining
contribution of $0.62$ coming from the transverse  helicity part.
In our model the ratio $[d\Gamma (D\rightarrow \rho \mu \nu_{\mu}
)/dq^{2}]    /[d\Gamma    (D\rightarrow   K^{*}   \mu   \nu_{\mu}
)/dq^{2}]_{m_{\mu}\neq 0}$ at this point ($0.087$) becomes $0.96$
compared to $0.97$ at  $q^{2}=0$ in the massless  limit.  So, the
$SU(3)_{F}$   limit  remains  steady  when  the  simple,   single
form-factor is replaced by a more complicated  collection of form
factors.  This  behaviour  breaks down only at higher  $q^2$ when
the effects of the  differing  masses of $\rho$ and $K^*$  become
obvious.  Figures  \ref{Fig2} and \ref{Fig3} show the effect
of a massive muon in $D\rightarrow \rho \mu \nu$.

\section{Rare   $B\rightarrow  K^{*}  \nu  \bar{\nu}  $
Decay}

The main reason for studying  the decay  $B\rightarrow  K^{*} \nu
\bar{\nu} $ is that in contrast to the decay $B\rightarrow  K^{*}
\ell  \bar{\ell}  $,  where  $\ell$  is  a  charged  lepton,  its
differential   decay  rate  does  not  have  any  singularity  at
$q^{2}=0$.  In  the  standard  model,   $B\rightarrow  K^{*}  \nu
\bar{\nu} $ decay is governed by $Z^{0}$ penguin diagrams and box
diagrams.  The decay $B\rightarrow K^{*} \ell \bar{\ell} $ has an
additional   structure    $\sigma_{\mu\nu}q_{\nu}/q^{2}$,   which
dominates  the decay  rate  \cite{OT91}.  This does not occur, of
course,  in those  calculations  that stay away from the  $q^2=0$
region.  Moreover, the decay $B\rightarrow  K^{*} \nu \bar{\nu} $
is a good  process  theoretically  , since both the  perturbative
$\alpha_{s}$ and  nonperturbative  $1/m^{2}_{b}$  corrections are
known to be small \cite{GrLiNa}.

Contributions  from the $Z^{0}$ penguin diagrams and box diagrams
are  sensitive  functions  of the top quark mass  $m_{t}$.  Thus,
they contain an  uncertainty  due to the dependence of $m_{t}$ on
the choice of the  renormalization  scale $\mu $.  As stressed in
ref.  \cite{BB},  in order  to  reduce  this  uncertainty,  it is
necessary  to  calculate  $O(\alpha_{s})$  corrections  to  these
diagrams  involving  internal top quark exchanges.  The resulting
effective  Hamiltonian  for  $B\rightarrow  K^{*} \nu \bar{\nu} $
decay is given \cite{BB} as follows:

\begin{eqnarray}   H^{\nu   \bar{\nu}}_{eff}   &   =  &
\frac{4G_{F}}{\sqrt{2}}\left(         \frac{\alpha}{2         \pi
\sin^{2}\theta_{W}}    \right)    V^{*}_{ts}    V_{tb}   X(x_{t})
(\bar{s}\gamma_{\mu  }  L   b)(\bar{\nu}\gamma_{\mu  }  L  \nu  )
\label{Heff} 
\end{eqnarray} 
where $x_{t}=m^{2}_{t}/M^{2}_{W}$ and
\begin{eqnarray} X(x) & = & X_{0}(x)+\frac{\alpha_{s}}{4 \pi } \,
X_{1}(x) 
\end{eqnarray}
Here, $X_{0}$ represents pure electroweak one-loop  contributions
and  $X_{1}$  results  from   $O(g^{4}_{2}\alpha_{s})$   two-loop
diagrams.  We  do  not  display  here  the   explicit   forms  of
$X_{0}(x_{t})$  and   $X_{1}(x_{t})$,   which  can  be  found  in
ref.\cite{BB}.  At $\mu =  M_{W}$  and  $m_{t}=175$  GeV, we find
that $X(x_{t})=1.47$.

The differential decay rate for $B\rightarrow K^{*} \nu \bar{\nu}
$ at zero  momentum  transfer is  
\begin{eqnarray}  
\frac{d\Gamma(B\rightarrow K^{*} \nu \bar{\nu})}{d q^{2}} \mid_{q^{2}=0} 
& = &
\frac{G^{2}_{F}}{192   \,  \pi^{3}   \,   m^{3}_{B}}   \,  \left(
\frac{\alpha }{2 \pi \sin^{2}\theta_{W}}  \right)^{2} |V^{*}_{ts}
V_{tb}|^{2} \, (m^{2}_{B}-m^{2}_{K^{*}})^{3}  \nonumber \\[0.4cm]
& \cdot &  |A^{B\rightarrow K^{*}}_{0}(0)|^{2} |X(x_{t})|^{2} 
 \label{eq:OT4}
\end{eqnarray}

Taking      $\tau_{B}=1.5       \times       10^{-12}$sec$^{-1}$,
$\sin^{2}\theta_{W}=0.23$,   $|V_{tb}|=1$  and   $A^{B\rightarrow
K^{*}}_{0}(0)=0.4$  \cite{OXT} and varying  $V_{ts}$ in the range
$0.030 \leq |V_{ts}|  \leq 0.048$ we find 
 
\begin{eqnarray}  
4.81 \times   10^{-8}  \,   GeV^{-2}   \;\;\;  \leq   \;\;\; 
\frac{1}{\Gamma_{Tot.}}
\frac{d \Gamma(B\rightarrow K^{*} \nu \bar{\nu}) }{dq^{2}} \;\;\; 
\leq \;\;\; 1.23 \times 10^{-7} \, GeV^{-2} 
\end{eqnarray}

The  differential   decay  rate  for   $B\rightarrow   \rho  \ell
\bar{\nu}_{\ell}$  is  determined  by the form factor  $A_{0}$ at
$q^{2}=0$, just as in the case of the $D$ decays.

The  ratio  of  $B\rightarrow  \rho  \ell   \bar{\nu}_{\ell}$  to
$B\rightarrow   K^{*}  \nu  \bar{\nu}$   gives  
 
\begin{eqnarray}
\frac{[d\Gamma    (B\rightarrow   \rho   \ell    \bar{\nu}_{\ell}
)/dq^{2}]_{q^{2} \rightarrow 0}}{[d\Gamma (B\rightarrow K^{*} \nu
\bar{\nu}    )/dq^{2}]_{q^{2}    \rightarrow    0}}    &    =   &
\frac{|V_{ub}|^{2}}{|V^{*}_{ts}V_{tb}|^{2}}
\left(\frac{m^{2}_{B}-m^{2}_{\rho}}{m^{2}_{B}-m^{2}_{K^{*}}}
\right)^{3}   \left(\frac{   2   \pi   \sin^{2}\theta_{W}}{\alpha
}\right)^{2} \nonumber \\[0.4cm]
& \cdot & \frac{|A^{B\rightarrow \rho}_{0}(0)|^{2}}{|A^{B\rightarrow
K^{*}}_{0}(0)|^{2}} \frac{1}{|X(x_{t})|^{2}}  \nonumber  \\[0.4cm] 
 & = & (1.86  \times  10^{4})  \,  \frac{|V_{ub}|^{2}}{|V^{*}_{ts}|^{2}}
\frac{|A^{B\rightarrow \rho}_{0}(0)|^{2}}{|A^{B\rightarrow 
K^{*}}_{0}(0)|^{2}} \label{eq:OT5} 
\end{eqnarray}

The   form    factors    $A^{B\rightarrow    \rho}_{0}(0)$    and
$A^{B\rightarrow  K^{*}}_{0}(0)$  have  already  been  calculated
\cite{OXT} \linebreak in the light- front quark model as shown in
the   table.   Thus,   we   see    \linebreak    $A^{B\rightarrow
\rho}_{0}(0)/ A^{B\rightarrow  K^{*}}_{0}(0)=0.75$.  That is, the
$SU(3)_{F}$  breaking of the form factors  becomes  larger as the
mass of the  decaying  meson  increases.  In ref.  \cite{LW}  the
double   ratio  of  form   factors   $(f^{(B\rightarrow\rho   )}/
f^{(B\rightarrow
K^{*})})/(f^{(D\rightarrow\rho)}/f^{(D\rightarrow  K^{*})})$, was
considered.  In our notation, this would correspond to the ratios
of the form factors  $A_{1}$.  They chose to write  everything in
terms of $A_{1}$ and ratios of the other form factors to $A_{1}$.
As  mentioned  above,  they  did  not  calculate  at  $q^2=0$,  a
kinematic  point at which  only one form  factor is needed.  This
double   ratio   should  be  equal  to  unity  in  the  limit  of
$SU(3)_{F}$.  However, explicit  calculation of form factors have
shown that  $(A_{0}^{(B\rightarrow\rho  )}/  A_{0}^{(B\rightarrow
K^{*})})/        (A_{0}^{(D\rightarrow\rho)}/A_{0}^{(D\rightarrow
K^{*})})  =0.85$,  i.e., an  $SU(3)_{F}$--breaking  effect at the
level of about 15 $\%$.  In ref.  \cite{LW} an argument  that the
$SU(3)$  symmetry  violation  could be small in the ratios of the
form  factors  and  that  a  determination   of  $|V_{ub}|$  with
theoretical  uncertainties of less than $10 \%$.  This may be the
case for the region of $y$ considered.  At $q^2 = 0$, however, in
taking the square of the double  ratios  the  ratios are  reduced
from the symmetry limit value of unity to $0.72$.

\section{Conclusion}

In this paper we have reviewed the present status of  theoretical
attempts to calculate the  semileptonic  charm and bottom decays.
We  then   presented  a  calculation   of  these  decays  in  the
light--front  frame at the kinematic point $q^2=0$.  This allowed
us to evaluate  the form factors at the same value of $q^2$, even
though the allowed  kinematic  ranges for charm and bottom decays
are very  different.  Also, at this kinematic  point the decay is
given in terms of only one form factor $A_{0}(0)$.  For the ratio
of the decay rates given by the E653  collaboration  we show that
the  determination  of the ratio of the CKM  matrix  elements  is
consistent with that obtained from the unitarity  constraint.  At
present, though, the unitarity method still has greater accuracy.
For $B$ decays, the decay $B \rightarrow  K^{*} \ell  \bar{\ell}$
at $q^2=0$  involves an extra form factor  coming from the photon
contribution and so is not amenable to the same kind of analysis,
leaving only the decay $B \rightarrow  K^{*} \nu  \bar{\nu}$ as a
possibility.  This   is   not   an   easy   mode   to   determine
experimentally.

The results obtained in our model for the form factor $A_{0}(0)$,
for $D$ decays, as well as other  models are  collected  in table
\ref{tab1}.  We see that theoretical  predictions of the ratio of
form-factors fall in a range near $1$.  If $A_{0}(0)$ is obtained
from Eq.  (\ref{def})  with the calculated  values of $A_{1}$ and
$A_{2}$  then there may be  difficulties  coming from the $q^{2}$
dependences   of  the  form   factors  and  also  from   possible
correlations in treating the errors.  The comparison with QCD sum
rules  predictions  \cite{Ball}  ,\cite{Nar},\cite{Slob}  shows a
similar problem (the exception is ref.  \cite{CFS}, where $A_{0}$
is  directly  calculated  ):  the  uncertainities  in the results
obtained using $A_{1}$ and $A_{2}$ are so large that they obscure
the real value of $A_{0}$.  For the  non--zero  lepton mass case,
use of $A_{1}$, $A_{2}$ and $V$ does not connect  smoothly to the
zero  lepton  mass  results.  When  $A_{2}$  is used in  place of
$A_{1}$ the $SU(3)$  symmetry  breaking remains small for a range
of $q^2$, even though there is a more  complicated  collection of
form factors.

It is interesting to note the predictions of \cite{BFO}  obtained
in a  framework  based on HQET and  chiral  symmetries.  Although
their  values for the form factors for  $D\rightarrow  \rho $ and
$D\rightarrow  K^{*} $ agree with the predictions of other models
given in the table, the result of  $B\rightarrow  \rho$ is larger
than most of the others.

It is claimed sometimes that the light-front quark model is ruled
out since it  typically  gives a value about $15\%$ less than one
for  the  ratio  $R=f^{D\rightarrow   \pi}_{+}(0)/f^{D\rightarrow
K}_{+}(0)$.  The  experimental  value of $R$ is obtained from the
measurements of the decays $D^{0}\rightarrow K^{-} e^{+} \nu_{e}$
and   $D^{+}\rightarrow   \pi^{0}  e^{+}  \nu_{e}$  by  MARK--III
\cite{MARK} and CLEO--II \cite{CLEO}:

\begin{eqnarray}
\frac{Br(D^{+}\rightarrow \pi^{0} e^{+} \nu_{e})}{Br(D^{0}\rightarrow 
K^{-} e^{+} \nu_{e})} & = & \left\{ \begin{array}{lll} 
(8.5 \pm 2.7 \pm 1.4 ) \; \% &  & MARK-III \\[0.3cm] 
(10.5 \pm 3.9 \pm 1.3 ) \; \% & & CLEO-II \end{array} \right. \nonumber
\end{eqnarray}
To  translate  these  results  into the values of ratio $R$, pole
dominance  is  assumed  for the  $q^{2}-$dependence  of the  form
factors for the $\pi e \nu_{e} (K e \nu_{e})$ decay with the mass
of  the   vector   resonance   given   by   the   mass   of   the
$D^{*}(D^{*}_{s})$ meson :
\begin{eqnarray}
R & = & \left\{\begin{array}{lll} 
(1.29 \pm 0.21\pm 0.11 ) & &  MARK-III \\[0.3cm] (1.01 \pm 0.20 \pm 0.07 ) 
& &  CLEO-II \end{array} \right. \nonumber
\end{eqnarray}
Given the size of these  errors, it is  premature to claim that a
value less than unity is ruled out.

In    an    analysis    of    two    body    hadronic     decays,
$D^+\rightarrow\pi^+\pi^0$  and  $D^0\rightarrow  K^+\pi^-$  Chau
et.al.  \cite{Cea}  calculated  the  ratio  $R$  and  found  that
relative  magnitude  of the  form  factors  should  be such  that
$f^{D\rightarrow  \pi}_{+}(0)>f^{D\rightarrow K}_{+}(0)$ in order
to be consistent with the pattern of $SU(3)$  breaking.  However,
this  calculation  relies  on  the  large-$N_{c}$   factorization
approach  in addition to the pole  dominance  assumption  for the
$q^{2}-$dependences  of the form  factors.  They also neglect the
final state interaction effects.  These assumptions have recently
been  questioned  by Kamal et.al.  \cite{Kamal}.  Also, the value
of the branching  ratio for  $D^0\rightarrow  K\pi$ may have been
overestimated \cite{ID}.

Finally,  we  note  that  as the  mass of the  decaying  particle
increases the $SU(3)$ symmetry breaking becomes greater at $q^2 =
0$.

\vspace{.3in}  
\centerline{ {\bf  Acknowledgment}} 

This  work was  supported  in part by the  Natural  Sciences  and
Engineering Council of Canada and by the Scientific and Technical
Research Council of Turkey.

\begin{table}
\baselineskip=20pt

\vspace{1.0cm}

\caption{The form factor $A_{0}(0)$ of $D\rightarrow
\rho$,    $D\rightarrow    K^{*}$,    $B\rightarrow   \rho$   and
$B\rightarrow K^{*}$ transitions.  \label{tab1}}

\begin{center}

\begin{tabular}{|c  | c| c| c| c| c|}  \hline\hline 
Reference & $D\rightarrow \rho $ & $D\rightarrow  K^{*}$ & $D\rightarrow \rho /
D\rightarrow  K^{*}$ &  $B\rightarrow  \rho $ &  $B\rightarrow K^{*}$ \\ \hline  
\cite{WSB}$^{a,\dag}$  & $0.67$ & $ 0.73$ & $0.92$ & $0.28$ & $-$ \\  
\cite{GIW}$^{a}$  & $0.85$  &  $0.80$ & $1.06$ & $0.14$ & $-$ \\ 
\cite{GilSin}$^{a}$  & $-$  &  $0.91 (0.84)$  &  $-$  &  $-0.37$  &  $-$  \\
\cite{FGM}$^{a}$ &  & $0.74 \pm 0.12$ & $-$ & $0.14 \pm 0.20$ & $-$ \\ 
\cite{Ball}$^{b}$  & $0.57 \pm 0.40$ & $0.45 \pm 0.30$  &  
$1.27   \pm   1.23$  &  $0.79   \pm   0.80$  &  $-$  \\
\cite{CFS}$^{b,\dag}$  & $0.52 \pm 0.05$ & $0.58 \pm 0.05$ & $0.90 \pm 0.12$ & 
$0.24 \pm 0.02$ & $0.30 \pm 0.03$ \\
\cite{Nar}$^{b}$ & $-$ & $-$ & $-$ & $0.15 \pm 0.97$ & $-$ \\ 
\cite{Slob}$^{b}$ & $-$ & $-$ & $-$ & $0.28 \pm 1.1 $ & $-$ \\
\cite{CDBFGN}$^{c}$  & $0.74$ & $0.59$ & $1.25$ & $0.24$ & $-$ \\ 
\cite{BFO}$^{c,\dag}$ & $0.73 \pm 0.17$ & $0.65 \pm  0.14$ & $1.12 \pm  0.36$ & 
$1.10 \pm  0.30$ & $-$ \\
\cite{HusIT}$^{c}$  & $-$ & $0.39 \pm  0.13$ & $-$ & $-$ & $-$ \\
\cite{Lub}$^{d}$ & $0.76 \pm 0.25$ & $0.72 \pm 0.17 $ & $1.06 \pm 0.43$ & $-$ & 
$-$ \\  
\cite{BElS}$^{d,\dag}$ & $0.64 \pm 0.17$ & $0.71 \pm 0.16$ & $0.90 \pm 0.31$ & 
$-$ & $-$ \\ 
\cite{Abada}$^{d}$  & $-$ & $0.77  \pm  0.29$ & $-$ & $-0.57  \pm 0.65$ & $-$ \\
\cite{UKQCD}$^{d,\dag}$    & $-$ & $-$ & $-$ & $(0.22-0.49)^{+13}_{\!\, -8}$ & 
$-$ \\
\cite{AmRo}$^{e}$ & $-$ & $0.48 \pm 0.12$ & $-$ & $-$ & $-$ \\
\cite{Jaus90}$^{f}$ & $-$ & $-$ & $-$ &  $-$ & $0.31$ \\  
\cite{Jaus96}$^{f}$ & $0.69$ & $0.78$ & $0.88$ &  $0.32$ & $-$ \\ 
\cite{OXT}$^{f,\dag}$  & $-$ & $-$ & $-$ & $0.30$ & $0.40$ \\
This Work$^{\dag}$ &  $0.66$  &  $0.75$  & $0.88$  & $-$ & $-$ \\ 
 \hline\hline
\end{tabular} 
\end{center}  

$^{a}$ Quark model \\ 
$^{b}$ QCD sum rules \\  
$^{c}$  HQET + chiral  perturbation  theory  \\  
$^{d}$ Lattice  calculation \\ 
$^{e}$ Heavy-quark-symmetry  \\
$^{f}$ Light-front  quark  model \\
$^{\dag}$ $A_{0}(0)$ is directly calculated.
\end{table}

\begin{table}

\vspace{1cm}

\caption{The first two columns show the partial helicity rates $d\Gamma_{(i)}/dq^{2}$, $i=(0),(+,-),
(t),(T)$ for  longitudinal, transverse, flip and total contributions 
at $q^{2}=0.087$ GeV$^{2}$ in units of $|V_{CKM}|^{2} 10^{10}$ sec$^{-1}$
GeV$^{-2}$. The last column gives the ratio. \label{tab2}}

\begin{center}
\begin{tabular}{|c  | c| c| c| }  \hline\hline 
$\frac{d\Gamma_{(i)}}{dq^{2}}$ & $D\rightarrow \rho $ & $D\rightarrow  K^{*}$ &
$D\rightarrow \rho / D\rightarrow  K^{*}$ \\[0.2cm] \hline  
$\frac{d\Gamma_{(0)}}{dq^{2}}$ & $4.05$ & $ 4.15$ & $0.97$  \\[0.2cm]  
$\frac{d\Gamma_{(+,-)}}{dq^{2}}$ & $0.62$ & $ 0.62$ & $1.00$  \\[0.2cm] 
$\frac{d\Gamma_{(t)}}{dq^{2}}$& $1.02$  &  $1.04 $  &  $0.98$    \\[0.2cm]
$\frac{d\Gamma_{(T)}}{dq^{2}}$ & $5.69$ & $5.81$ & $0.97$  \\ \hline\hline
\end{tabular} 
\end{center}  
\end{table}

\newpage

\begin{center}

Figure Captions

\end{center}

\end{document}